\documentclass[3p]{elsarticle}
\UseRawInputEncoding
\usepackage{lineno,hyperref}
\usepackage{graphicx}
\usepackage{amsmath}
\usepackage{color}
\usepackage[normalem]{ulem}

\journal{arXiv}
\begin{document}

\begin{frontmatter}

\title{Asymmetry in Distributions of Accumulated Gains and Losses in Stock Returns}

\author[mymainaddress]{Hamed Farahani}
\author[mymainaddress]{R. A. Serota\fnref{myfootnote}}
\fntext[myfootnote]{serota@ucmail.uc.edu}

\address[mymainaddress]{Department of Physics, University of Cincinnati, Cincinnati, Ohio 45221-0011}

\begin{abstract}
We study decades-long historic distributions of accumulated S\&P500 returns, from daily returns to those over several weeks. The time series of the returns emphasize major upheavals in the markets -- Black Monday, Tech Bubble, Financial Crisis and Covid Pandemic -- which are reflected in the tail ends of the distributions. De-trending the overall gain, we concentrate on comparing distributions of gains and losses. Specifically, we compare the tails of the distributions, which are believed to exhibit power-law behavior and possibly contain outliers. Towards this end we find confidence intervals of the linear fits of the tails of the complementary cumulative distribution functions on a log-log scale, as well as conduct a statistical U-test in order to detect outliers. We also study probability density functions of the full distributions of the returns with the emphasis on their asymmetry. The key empirical observations are that the mean of de-trended distributions increases near-linearly with the number of days of accumulation while the overall skew is negative -- consistent with the heavier tails of losses -- and depends little on the number of days of accumulation. At the same time the variance of the distributions exhibits near-perfect linear dependence on the number of days of accumulation, that is it remains constant if scaled to the latter. Finally, we discuss the theoretical framework for understanding accumulated returns. Our main conclusion is that the current state of theory, which predicts symmetric or near-symmetric distributions of returns cannot explain the aggregate of  empirical results.
\end{abstract}


\end{frontmatter}

\section{Introduction \label{intro}}


Research on asymmetry of stock returns has a long and storied history \cite{french1987expected,campbell1992asymmetric,glosten1993relation,duffee1995stock,braun1995good,bekaert2000asymmetry,cont2001empirical,wu2001determinants,hong2003differences,zaluskakotur2006comparison,sive2009temporal,chakraborti2011econophysics,albuquerque2012skewnwss,sandor2016timescale,neuberger2021skewness,lee2023estimating}. Clearly there are many aspects of asymmetry and approaches to study it, such as the first passage time \cite{zaluskakotur2006comparison, sive2009temporal, sandor2016timescale}, differences between firm-level and overall market performance \cite{duffee1995stock, braun1995good, bekaert2000asymmetry, albuquerque2012skewnwss}, and many others. The simplest form of asymmetry is that overall there is a considerable gain in the stock market: financial advisors like to tell their clients that, on average, there is roughly a 10\% annual gain or, more precisely, 12\% gain -- see straight line in Fig. \ref{trend} -- minus 2\% average inflation. Of course, there are periods of market stagnation, decline and rapid growth -- such fluctuations around the overall growth trend are attributed to market volatility.

A far more interesting question is the asymmetry between gains and losses once the overall growth trends is already accounted for, that is when the data is de-trended. In this regard, of the otherwise numerous empirical properties of the raw market data \cite{cont2001empirical,chakraborti2011econophysics}, our interest is centered mainly on asymmetry as related to heavy tails of the distributions of gains and losses. Towards this end we study distributions of stock returns of the S\&P500 index, from daily returns to those accumulated over longer periods of time. Namely, we perform linear fits (LF) of the tails of complementary cumulative distribution functions (CCDF) of gains and losses on log-log scale to test for their power-law dependence. We compute confidence intervals (CI) \cite{janczura2012black} of LF, as well as conduct a statistical U-test \cite{pisarenko2012robust} in order to test for possible outliers, such as Dragon Kings \cite{sornette2012dragon} (DK) and negative Dragon Kings (nDK) \cite{pisarenko2012robust}.


We also perform numerical measures of the full distributions of returns using their probability density functions (PDF). Namely, we evaluate dependence on the number of days of accumulation of the mean, variance, Fisher-Pearson coefficient of skewness and first and second Pearson coefficients of skewness. The key results from those measures are that the mean of de-trended distributions increases near-linearly with the number of days of accumulation while the overall skew is negative -- consistent with the heavier tails of losses observed from PDF and CCDF -- and depends little on the number of days of accumulation. At the same time the variance of the distributions exhibits near-perfect linear dependence on the number of days of accumulation. While the near-linear shift of the mean can be easily accounted for phenomenologically, the current state of theory based on continuous stochastic differential equations (SDE) does not properly describe statistical measures of the distributions, especially skewness.

This paper is organized as follows. In Sec. \ref{initial} we explain de-trending procedure of returns, present time series of returns for years 1980-2024, and discuss the number of days of gains and losses - all in terms of the number of days of accumulation. In Sec. \ref{ccdf} we compare distributions of gains and losses using CCDF, including LF statistical tests for outliers. In Sec. \ref{full} we study full distributions of returns and their statistical measures: mean, variance and skewness. In Sec. \ref{theory} we address the state of theory vis-a-vis our empirical observations. Sec. \ref{conclusions} summarizes our main results.

\section{Empirical Results \label{empirical}}

\subsection{Initial Analysis of Returns \label{initial}}

\begin{figure}[!htb]
    \centering
    \includegraphics[width=1\linewidth]{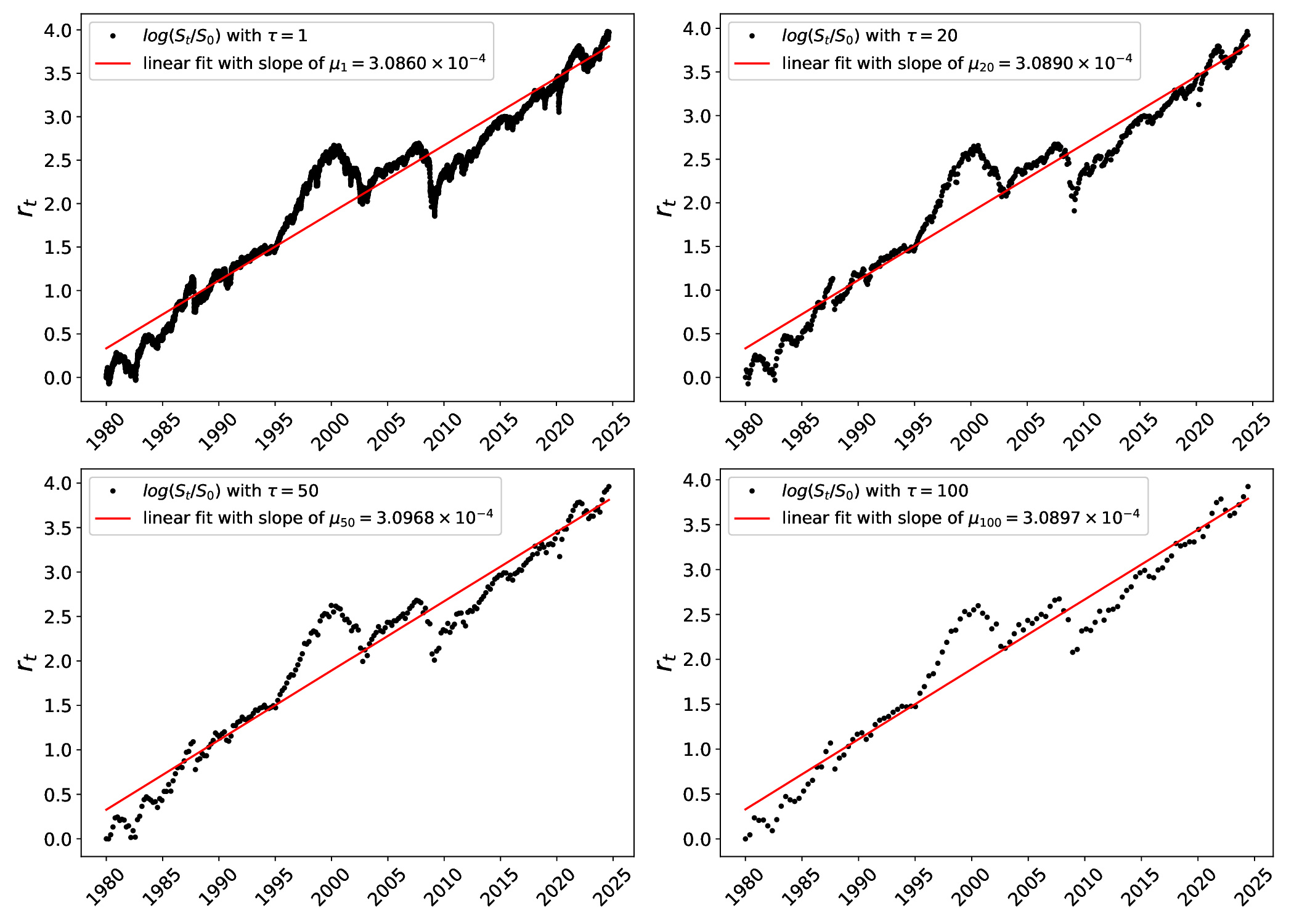}
    \caption{Linear fits of $r_{t} = \log\left(S_t/S_0\right)$ for $t=n \tau, n=0, 1, ...$, with $\tau=1, 20, 50, 100$ respectively.} 
    \label{trend}
\end{figure}

\begin{figure}[!htb]
    \centering
    \includegraphics[width=.6\linewidth]{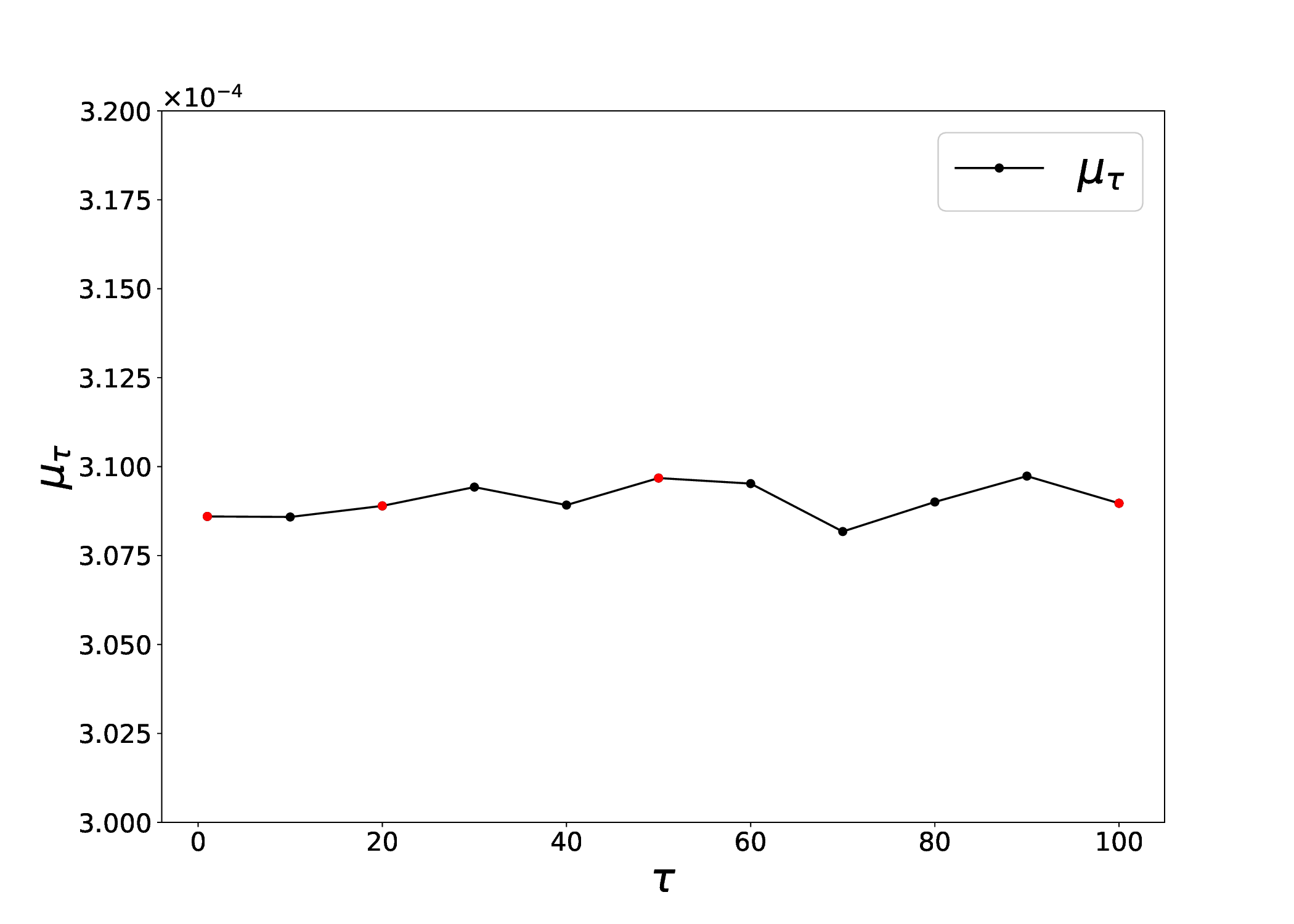}
    \caption{Slopes of linear fits of log returns $r_t$ for $t=n \tau, n=0, 1, ...$, as a function of $\tau$.}
    \label{mu_tau}
\end{figure}

With $S_t$ being the stock price, the linear upward trend of log returns

\begin{equation}
r_{t} = \log\left(\frac{S_t}{S_0}\right)
\label{logpriceratio}
\end{equation}

is shown in Fig. \ref {trend} for $t=n \tau, n=0, 1, ...$with $\tau=1, 20, 50, 100$ and the plot of slopes $\mu_{\tau}$ shown in Fig. \ref{mu_tau}.

De-trended log returns (or simply "returns" below) accumulated over time period $\tau$ are then given by 
\begin{equation}
dx_t = x_{t +\tau} - x_{t }= r_{t+\tau} - r_{t} - \mu \tau =  \log\left(\frac{S_{t+\tau}}{S_t}\right) - \mu \tau
\label{dx_t}
\end{equation}
\emph{where from now on we slide $t$ by one day when obtaining distributions as a function of $\tau$ and thus use $\mu=\mu_{1}$ -- the slope of daily log returns}, although clearly $\mu_\tau$ shows only very insignificant dependence on $\tau$.

\begin{figure}[!htb]
    \centering
    \includegraphics[width=1.1\linewidth]{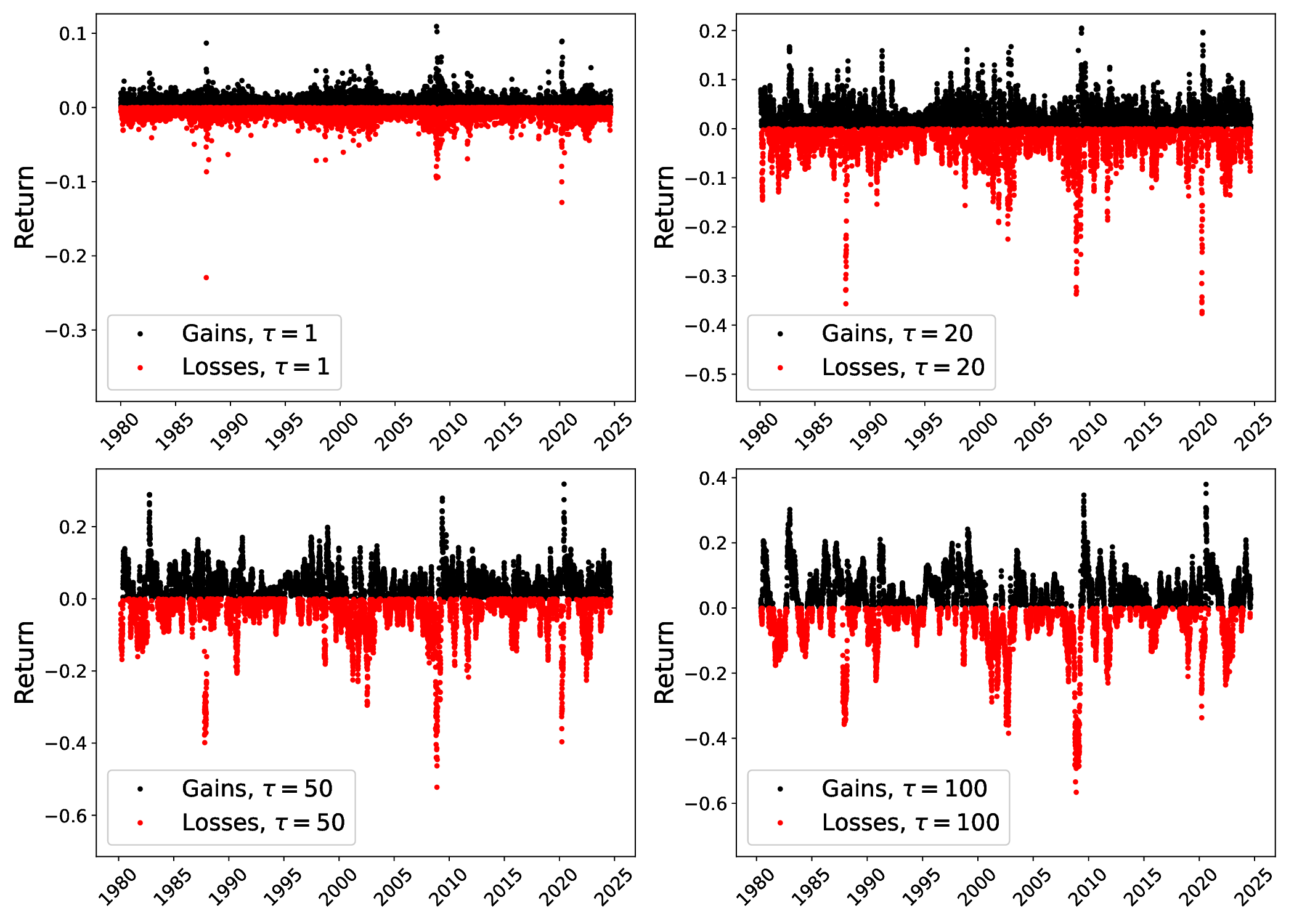}
    \caption{Time series of daily returns, $\tau=1$, and accumulated returns for $\tau=20, 50, 100$.}
    \label{timeseries}
\end{figure}

Fig. \ref{timeseries} shows time series of returns from 1980 to 2024. Notice obvious similarity with time series of realized volatility \cite{liu2023dragon}. Clearly, the largest negative peaks occurred during Black Monday, Tech Bubble, Financial Crisis and Covid Pandemic. Not surprisingly, following those drops, the largest positive peaks occurred relatively shortly after. 

\begin{figure}[!htb]
    \centering
    \includegraphics[width=1\linewidth]{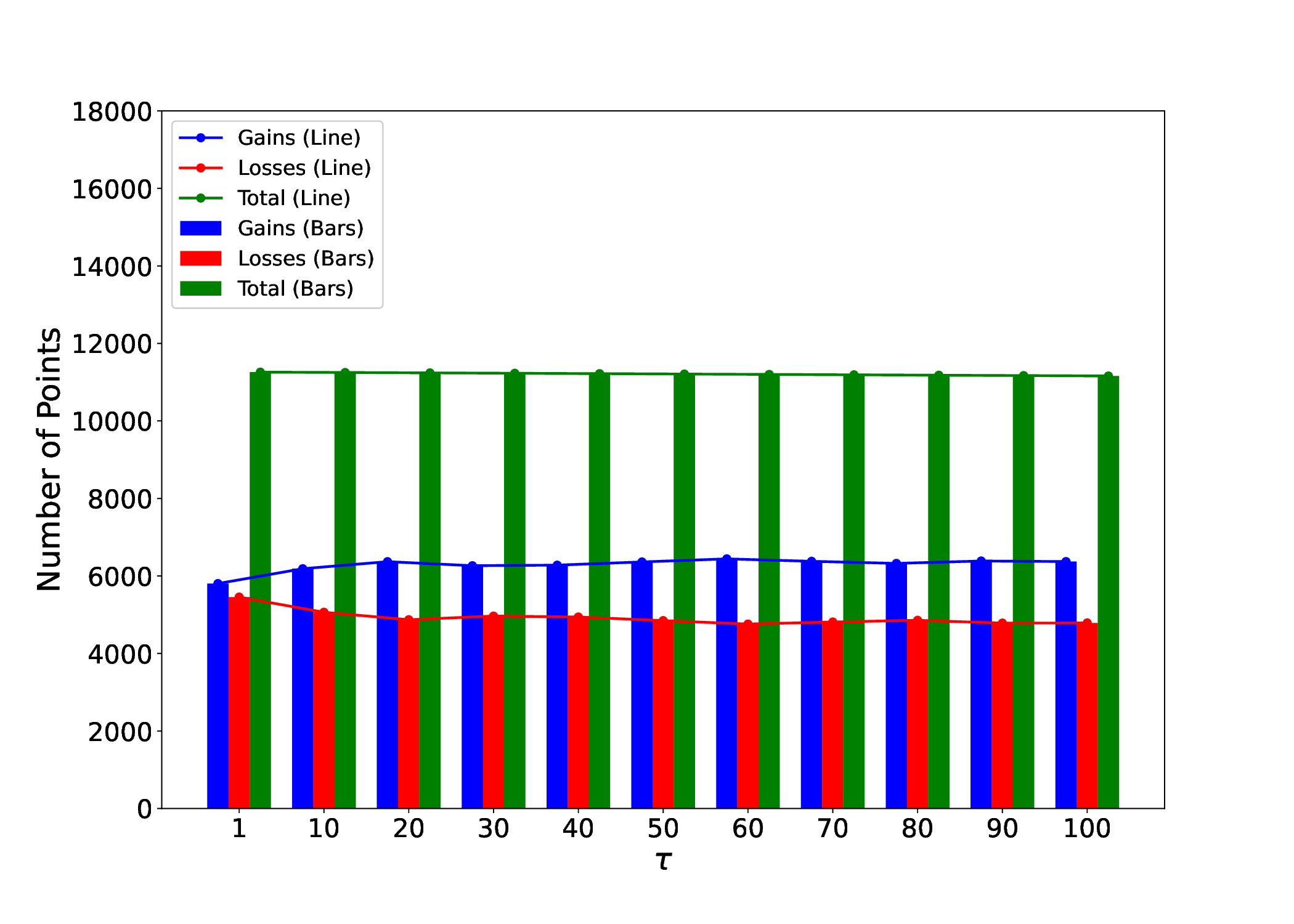}
    \caption{Number of data points for gains and losses and their sum (the total number of points) as a function of $\tau$.}
    \label{points_gains-losses}
\end{figure}

Fig. \ref{points_gains-losses} shows the number of data points for gains and losses as a functions of $\tau$, as well as their sum -- the total number of points in the data set -- for the same 1980-2024 time period as the time series in Fig. \ref{timeseries}. For illustrative purposes, the numbers are explicitly shown for $\tau=1, 5, 10, 20$ in Table \ref{tab_points_gains-losses}. Clearly, the number of gains increases as a function of $\tau$ while the number of losses decreases. The total number of points is given by $11259 - \tau +1$, where $11259$ is the size of the data set for daily returns, $\tau = 1$.

\begin{table}[!htb]
    \centering
    \begin{tabular}{|c|c|c|c|}
        \hline
        \textbf{$\tau$} & \textbf{Total Points} & \textbf{Losses} & \textbf{Gains} \\
        \hline
        1 & 11259 & 5455 & 5804 \\
        5 & 11255 & 5167 & 6088 \\
        10 & 11250 & 5063 & 6187 \\
        20 & 11240 & 4871 & 6369 \\
        \hline
    \end{tabular}
    \caption{Summary of Total Points, Losses, and Gains for Different $\tau$ Values}
    \label{tab_points_gains-losses}
\end{table}

\subsection{Distributions of Gains and Losses \label{ccdf}}

\begin{figure}[!htb]
    \centering
    \includegraphics[width=.77\linewidth]{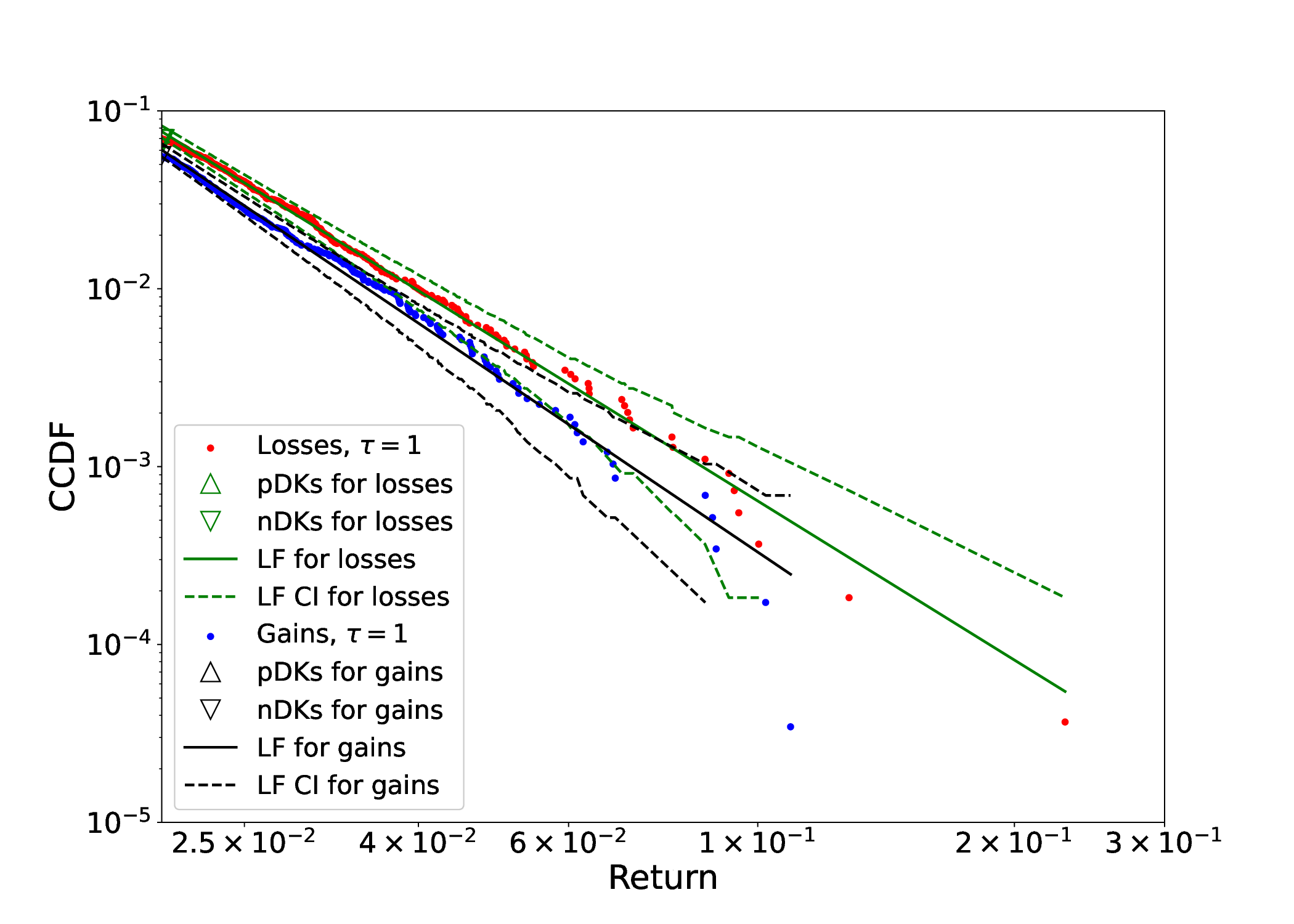}
    \caption{Linear fits of CCDF tails for gains and losses of daily returns, $\tau=1$, with CI (dashed lines) and possible DK (pDK), denoted by up triangles, and negative DK (nDK) denoted by down triangles} .
    \label{CCDF1}
\end{figure}

\begin{figure}[!htb]
    \centering
    \includegraphics[width=.77\linewidth]{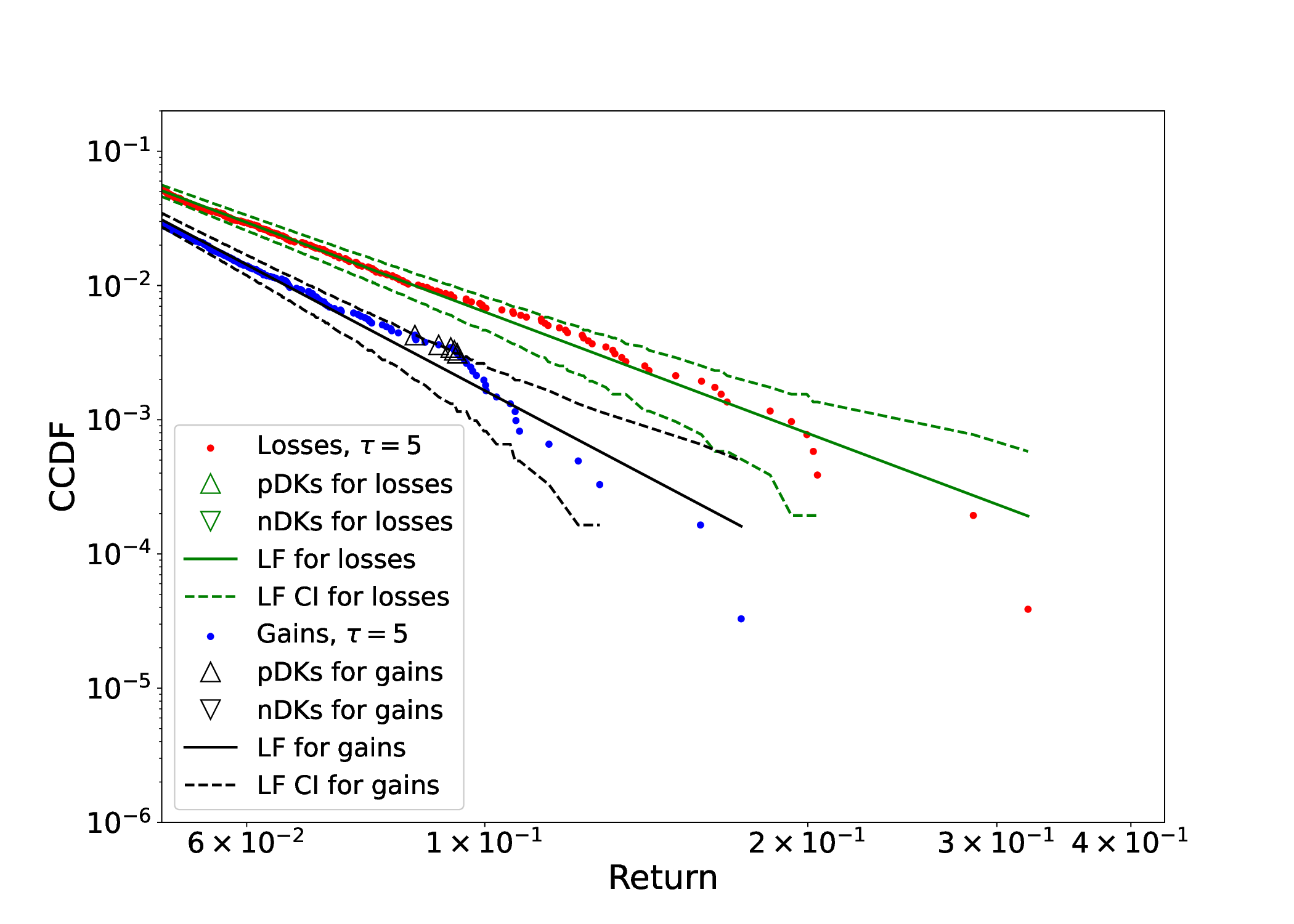}
    \caption{Linear fits of CCDF tails for gains and losses of $\tau=5$ accumulated returns with CI (dashed lines) and possible DK (pDK), denoted by up triangles, and negative DK (nDK) denoted by down triangles}
    \label{CCDF5}
\end{figure}

\begin{figure}[!htb]
    \centering
    \includegraphics[width=.77\linewidth]{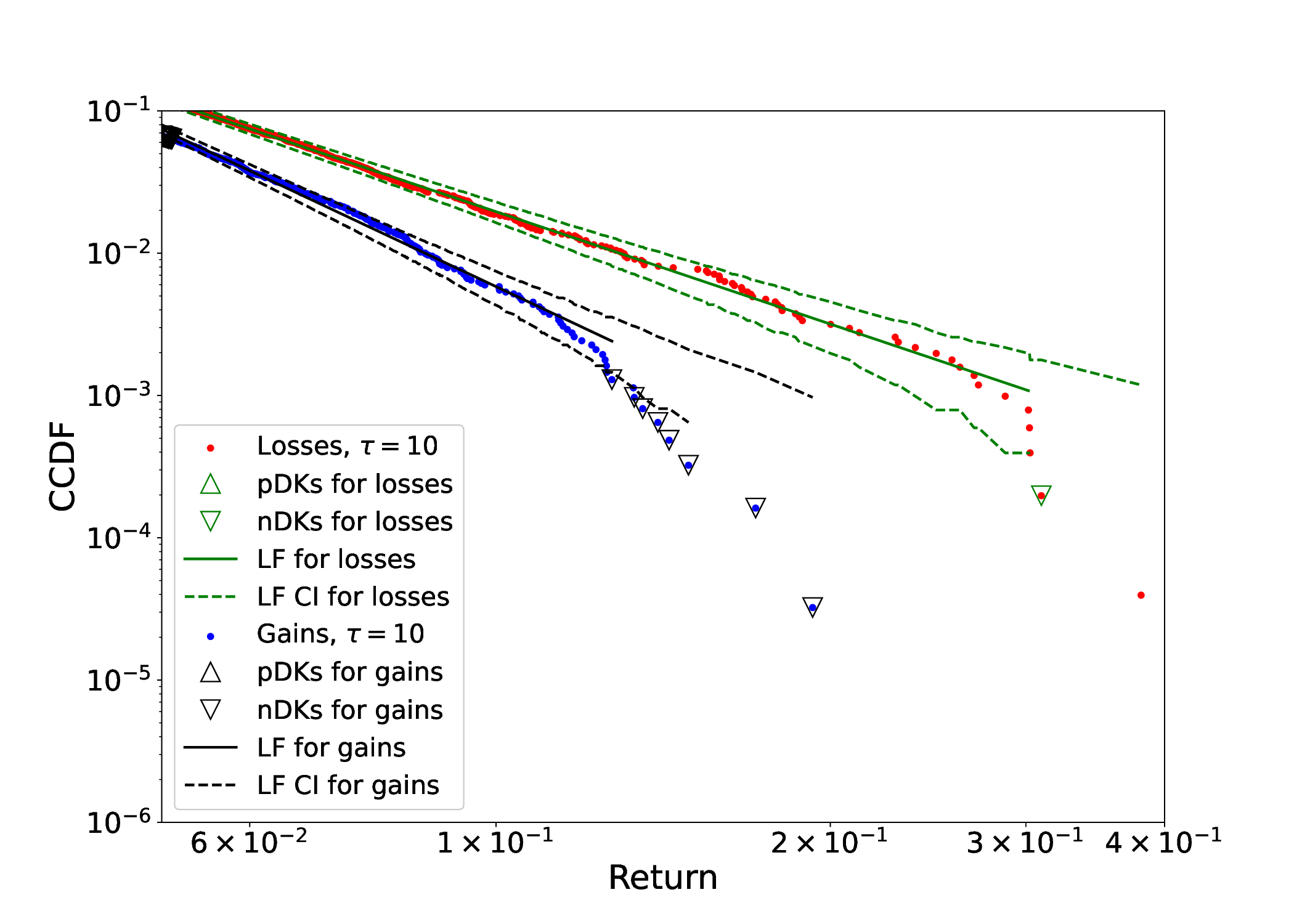}
    \caption{Linear fits of CCDF tails for gains and losses of $\tau=10$ accumulated returns with CI (dashed lines) and possible DK (pDK), denoted by up triangles, and negative DK (nDK) denoted by down triangles}
    \label{CCDF10}
\end{figure}

\begin{figure}[!htb]
    \centering
    \includegraphics[width=.77\linewidth]{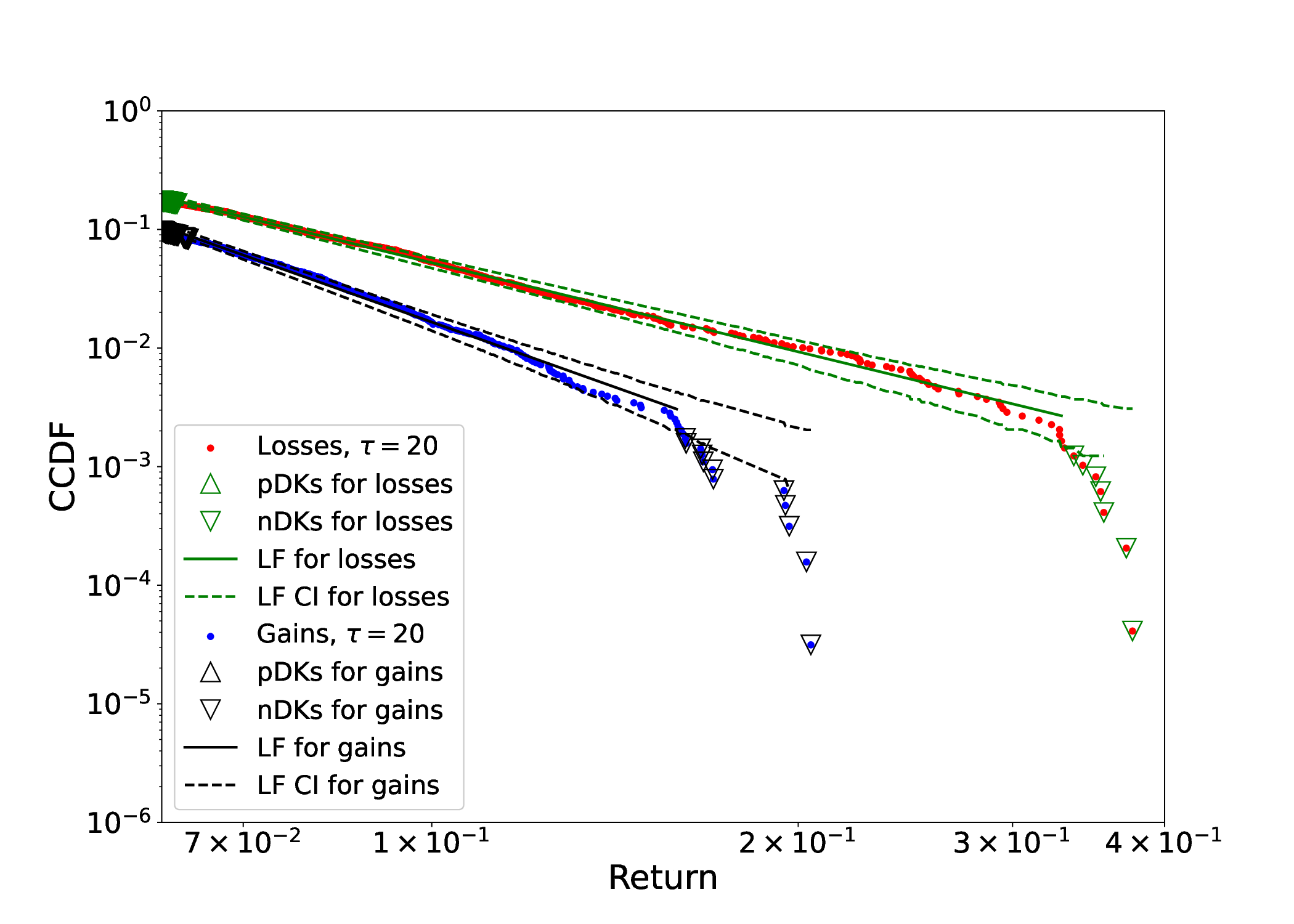}
    \caption{Linear fits of CCDF tails for gains and losses of $\tau=20$ accumulated returns with CI (dashed lines) and possible DK (pDK), denoted by up triangles, and negative DK (nDK) denoted by down triangles}
    \label{CCDF20}
\end{figure}

Figs. \ref{CCDF1} - \ref{CCDF20} show CCDF,  $1 - F_g (x)$ and $1 - F_l (x)$, of gains and losses on a log-log scale for $\tau=1, 5, 10, 20$. Here 
\begin{equation}
	\begin {split}
	F_g (x) = \int_{-\infty}^{x} f(x) \mathrm{d}x \\
	F_l (x) = \int_{\infty}^{x} f(x) \mathrm{d}x 
	\end{split}
\label{CCDF}
\end{equation}
are CDF of gains and losses respectively and  $f(x)$ is the PDF of returns (see Figs. \ref{PDF1} - \ref{PDF100} below). Also shown are linear fits of the tails, including their confidence intervals (CI) \cite{janczura2012black} and the results of the U-Test \cite{pisarenko2012robust} to identify outliers, such as DK and nDK. While daily returns seem to exhibit rather well-defined linear dependence, for larger $\tau$ the tail behavior is more complex with what might be called a developing shoulder and rapid drop-offs at tail ends. In this regard instead of thinking of possible DK (pDK) and nDK perhaps up and down triangles obtained from U-test and crossing lines of CI can be simply an indicator of poor goodness of fit. Again, notice obvious similarities with the tail behavior of realized volatility \cite{liu2023dragon}. 

\subsection{Full Distributions of Returns and their Statistical Measures \label{full}}

\begin{figure}[!htb]
    \centering
    \includegraphics[width=.77\linewidth]{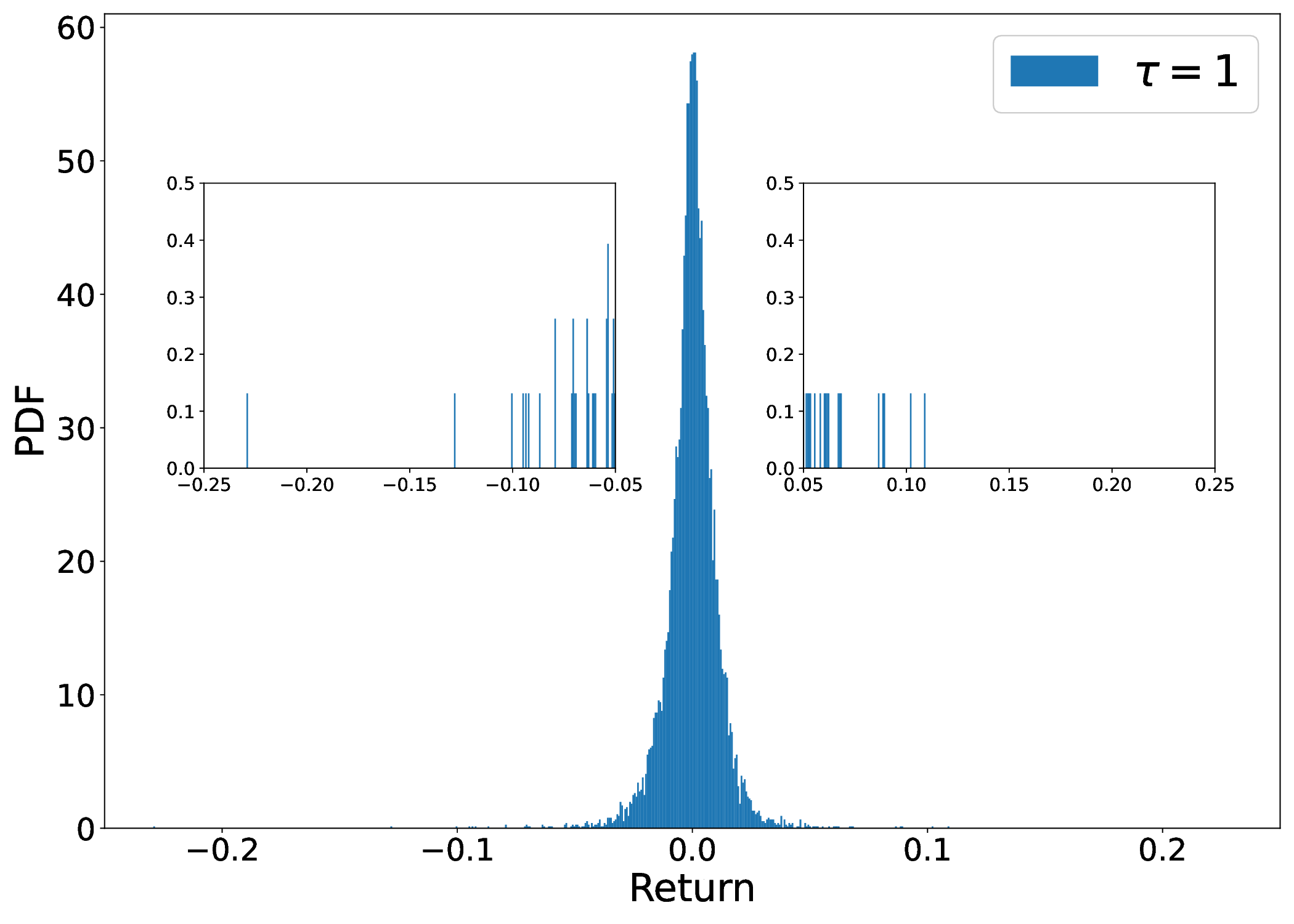}
    \caption{PDF of daily returns, with inserts showing tails of the distribution}
    \label{PDF1}
\end{figure}
\begin{figure}[!htb]
    \centering
    \includegraphics[width=.77\linewidth]{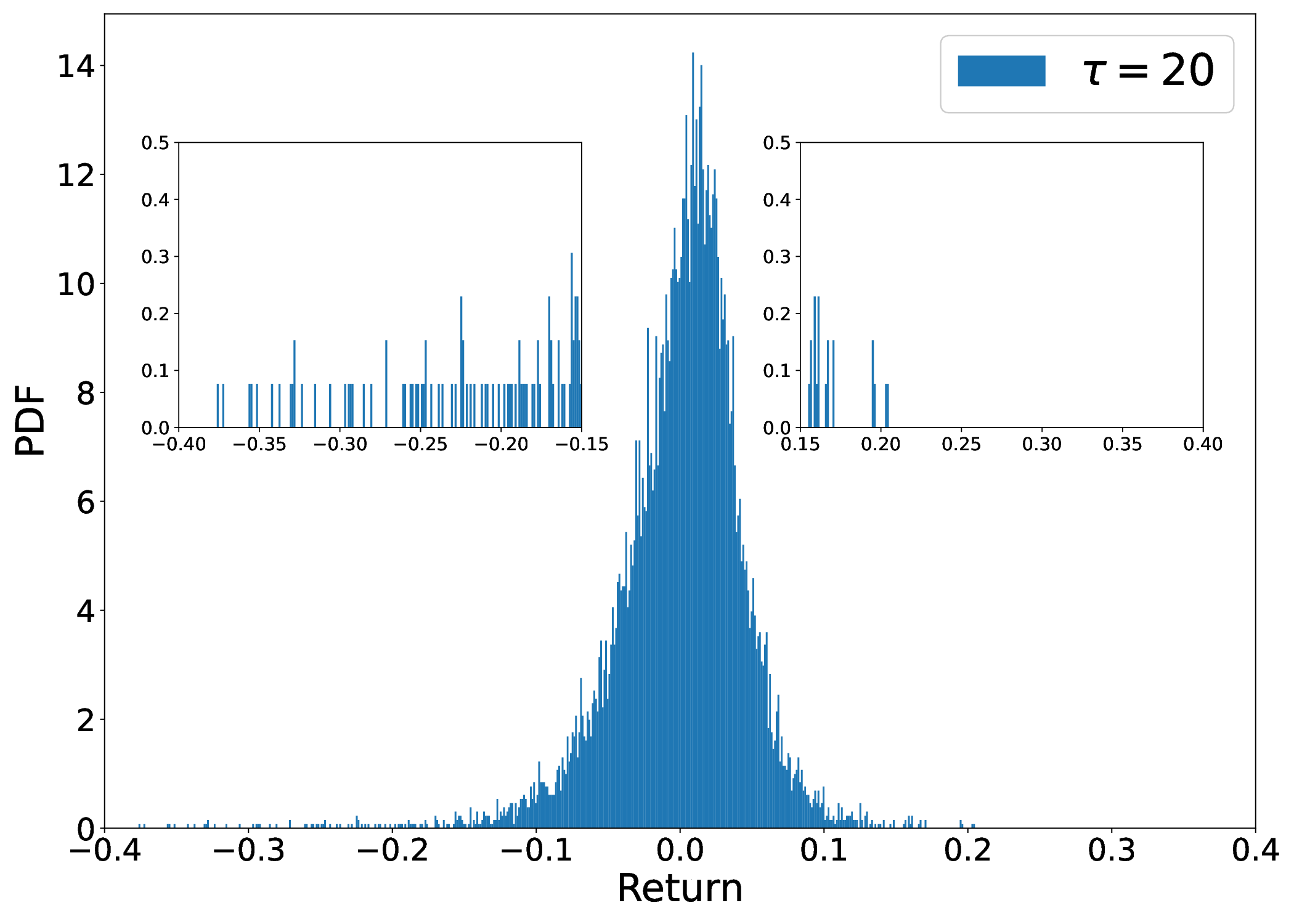}
    \caption{PDF of $\tau=20$ accumulated returns, with inserts showing tails of the distribution}
    \label{PDF20}
\end{figure}
\begin{figure}[!htb]
    \centering
    \includegraphics[width=.77\linewidth]{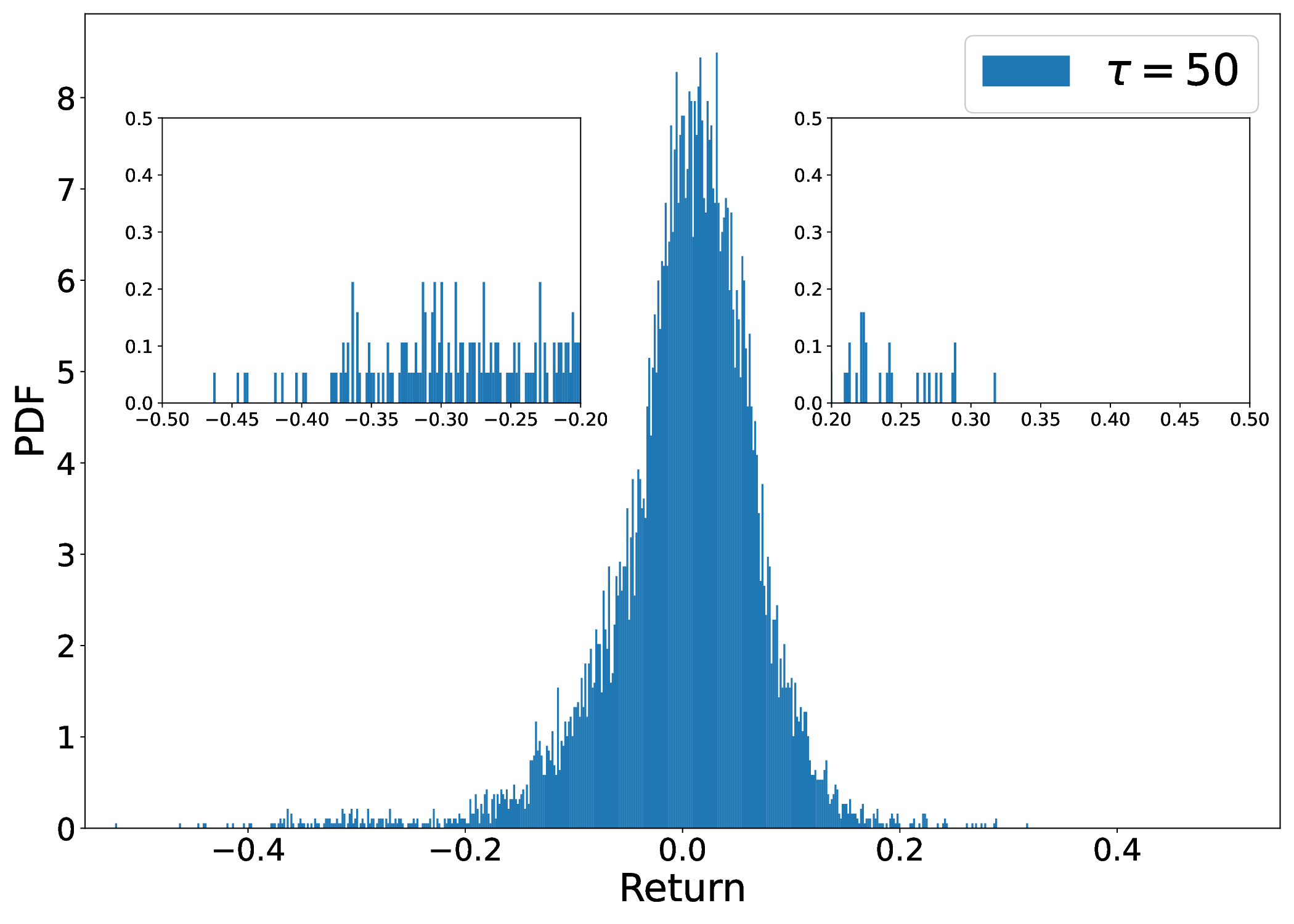}
    \caption{$\tau=50$ accumulated returns, with inserts showing tails of the distribution}
    \label{PDF50}
\end{figure}
\begin{figure}[!htb]
    \centering
    \includegraphics[width=.77\linewidth]{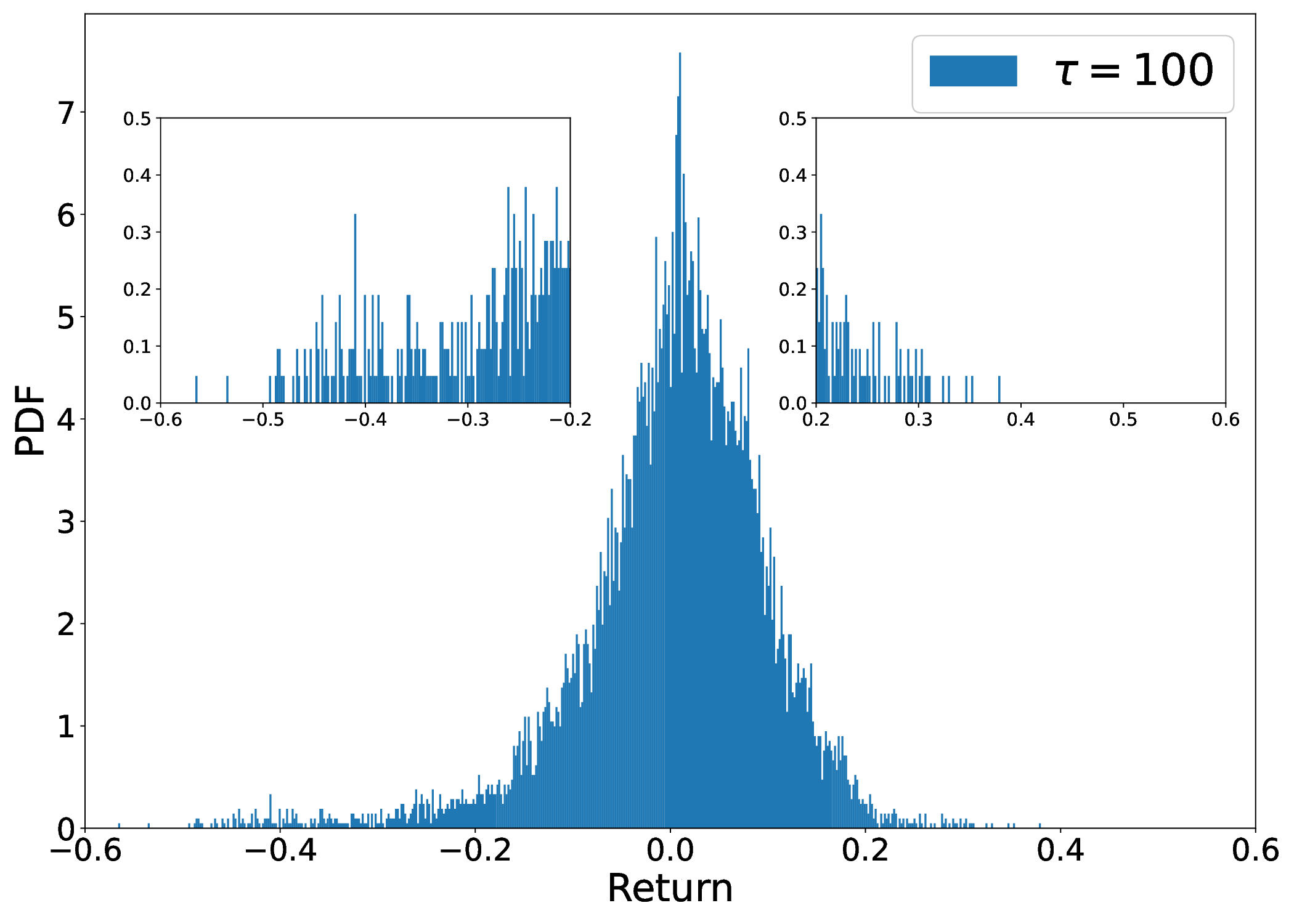}
    \caption{$\tau=100$ accumulated returns, with inserts showing tails of the distribution}
    \label{PDF100}
\end{figure}

Figs. \ref{PDF1} - \ref{PDF100} show PDF of daily and accumulated returns for $\tau=1, 20, 50, 100$. Clearly, PDF exhibit asymmetry and longer tails for losses versus gains \cite{palomar2018financial}, which are becoming more pronounced with larger $\tau$.

Next, we address mean, $m_1$, variance, $m_2$, and skewness of distributions in Figs. \ref{PDF1} - \ref{PDF100}. For the latter we employ Fisher-Pearson coefficient of skewness and first and second Pearson coefficients of skewness, defined respectively by 
\begin{equation}
\zeta=\frac{m_3}{m_2^{3/2}}
\label{zeta}
\end{equation}
\begin{equation}
\zeta_1=\frac{\left(m_1-\overline{m}\right)}{m_2^{1/2}}
\label{zeta1}
\end{equation}
\begin{equation}
\zeta_2=\frac{3\left(m_1-\widetilde{m}\right)}{m_2^{1/2}}
\label{zeta2}
\end{equation}
where $m_3$ is the third central moment of the distribution, $m_2$ is the variance, $m_2^{1/2}$ is the standard deviation, $\overline{m}$ is the mode, and $\widetilde{m}$ is the median. The reason for using $\zeta_1$ and $\zeta_2$ is that mathematically $m_3$ does not exist for power-law tails with slopes in Figs. \ref{CCDF1} -  \ref{CCDF20} as per linear fits (LF). These slopes are listed in Table \ref{slopes}, with PDF tail slopes being $+1$ of those for CCDF.

\begin{table}[!htb]
    \centering
    \begin{tabular}{|c|c|c|}
        \hline
        \textbf{$\tau$} & \textbf{Slope of Losses} & \textbf{Slope of Gains} \\
        \hline
        1 & -2.971 & -3.234 \\
        5 & -3.003 & -4.228 \\
        10 & -2.623 & -3.673 \\
        20 & -2.494 & -3.647 \\
        \hline
    \end{tabular}
    \caption{Slopes of linear fits of CCDF tails of losses and gains in Figs. \ref{CCDF1} -  \ref{CCDF20} as a function of $\tau$}
    \label{slopes}
\end{table}

\begin{figure}[!htb]
    \centering
    \includegraphics[width=.77\linewidth]{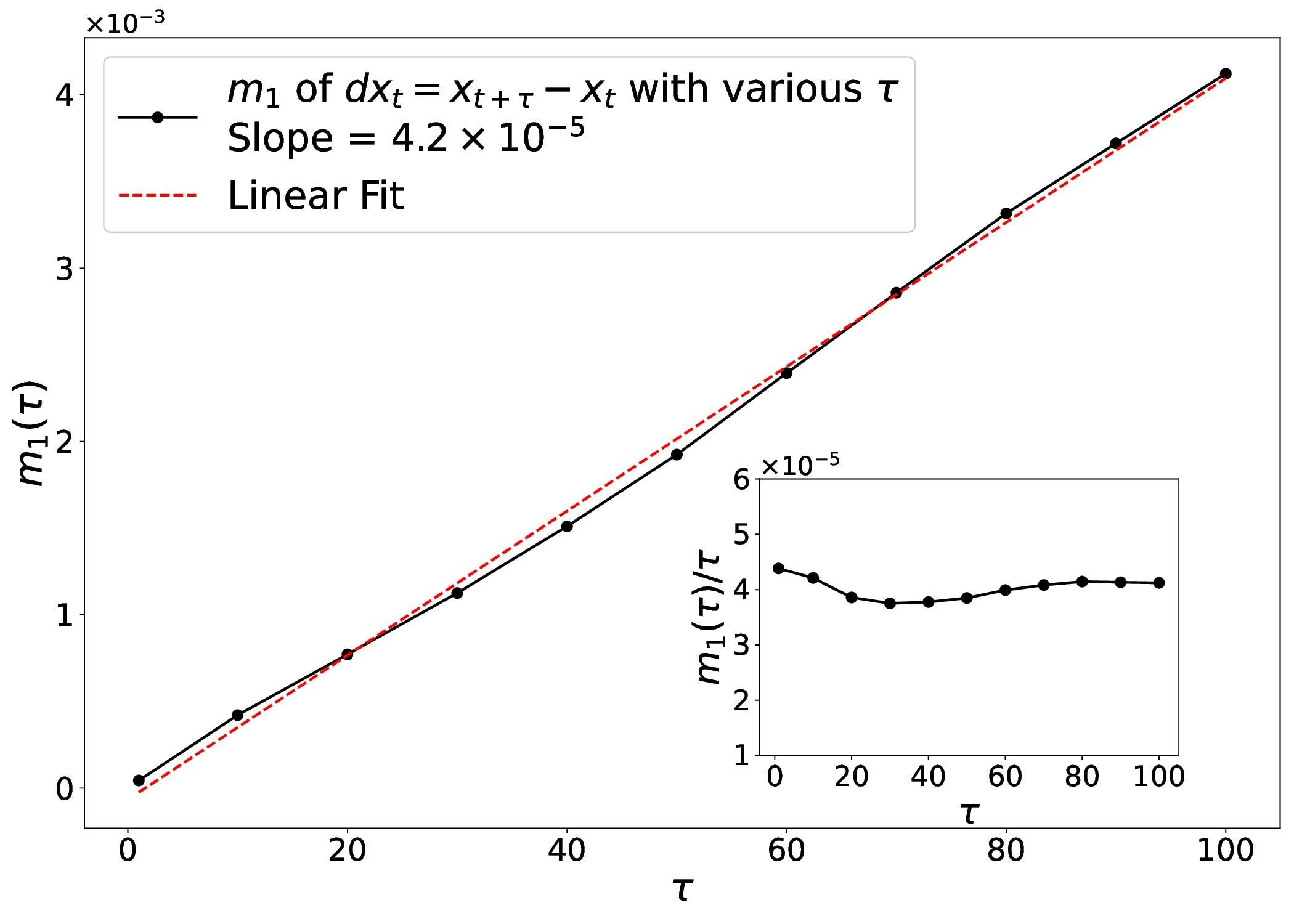}
    \caption{Mean of the distribution of returns as a function of the number of days of accumulation, $m_1(\tau)$, with its linear fit. Insert shows scaled mean $m_1(\tau)/\tau$.}
    \label{m1}
\end{figure}

\begin{figure}[!htb]
    \centering
    \includegraphics[width=.77\linewidth]{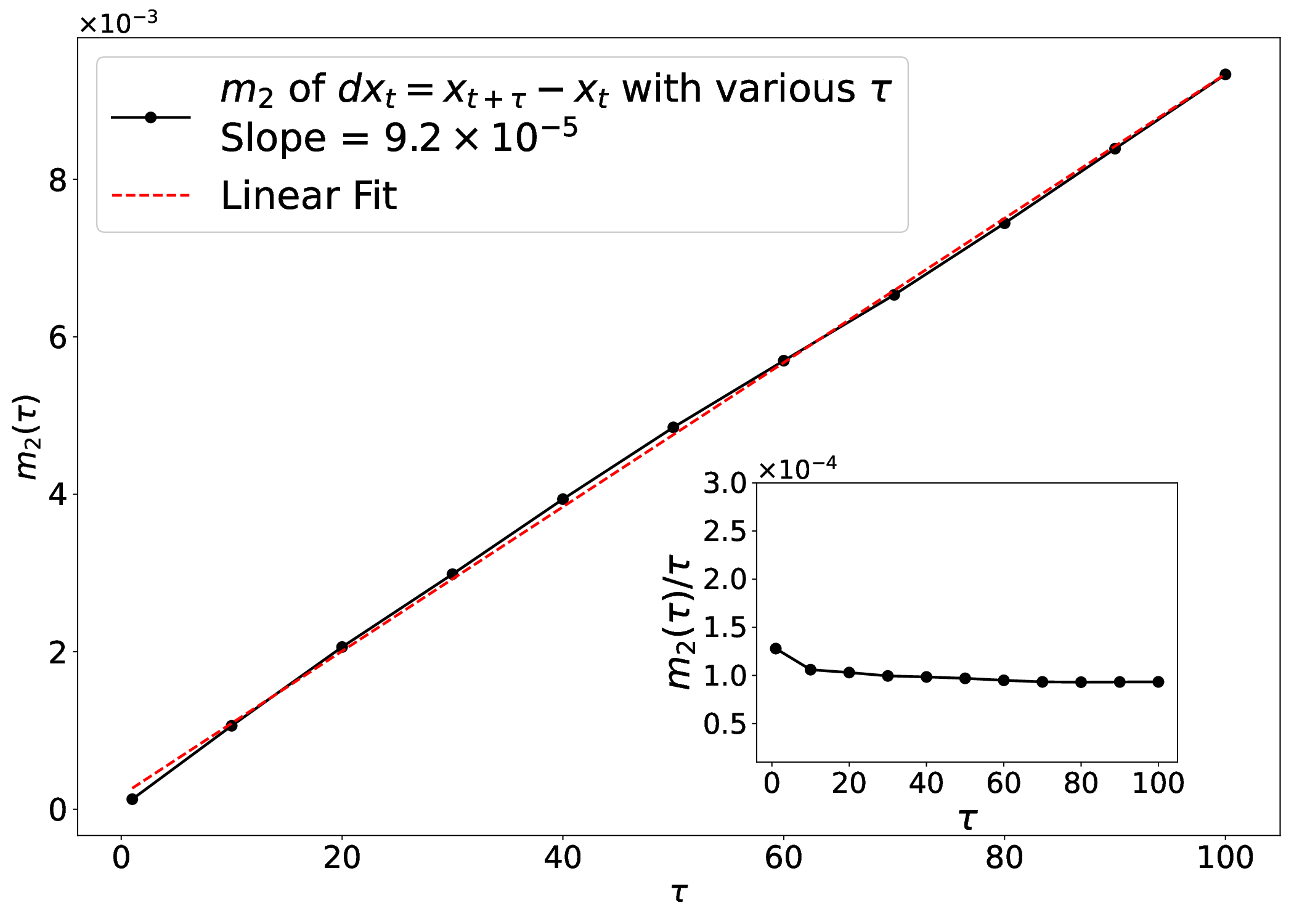}
    \caption{Variance of the distribution of returns as a function of the number of days of accumulation, $m_2(\tau)$, with its linear fit. Insert shows scaled variance $m_2(\tau)/\tau$.}
    \label{m2}
\end{figure}

\begin{figure}[!htb]
    \centering
    \includegraphics[width=.77\linewidth]{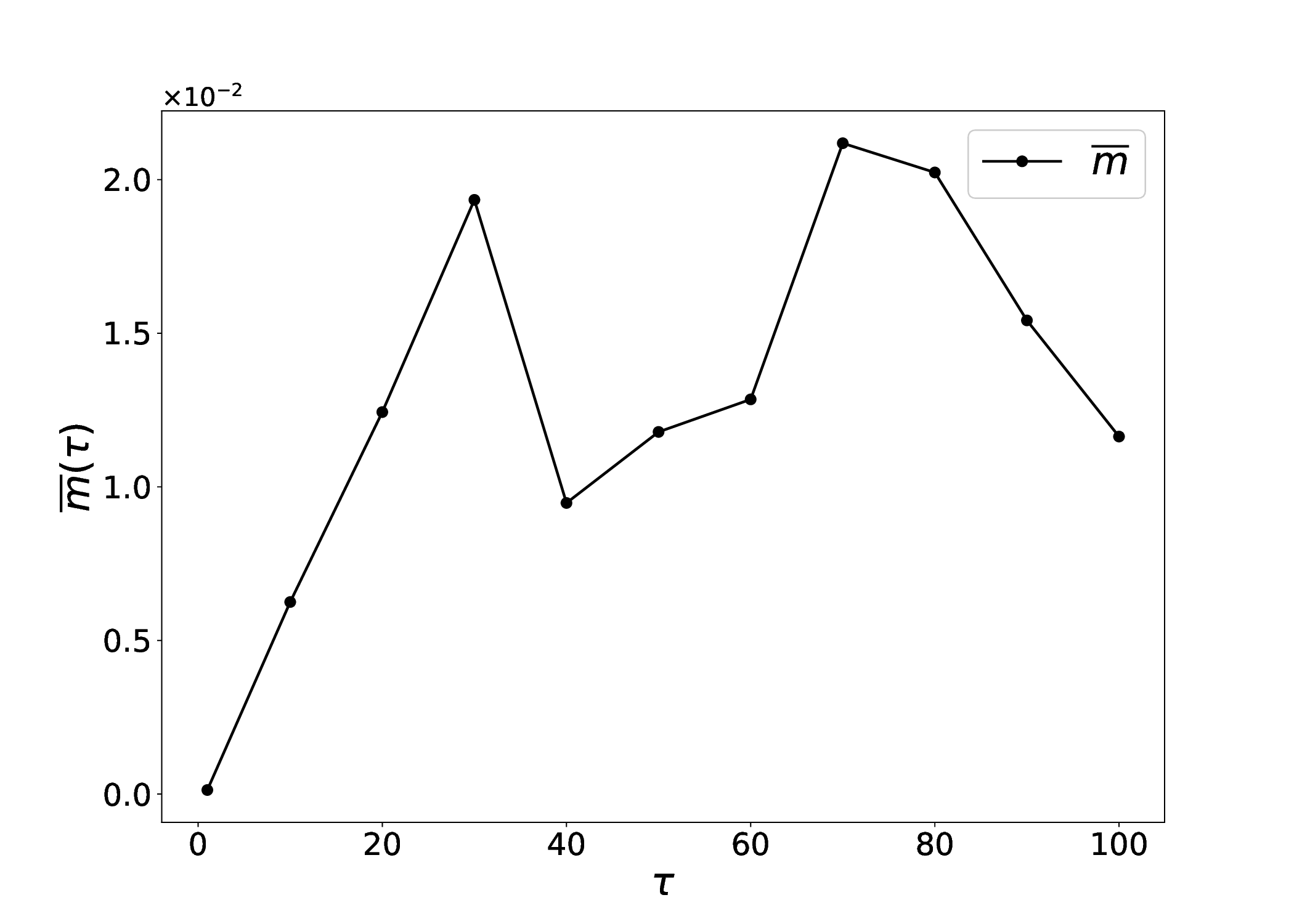}
    \caption{Mode of the distribution of returns as a function of the number of days of accumulation, $\overline{m}(\tau)$}
    \label{mbar}
\end{figure}

\begin{figure}[!htb]
    \centering
    \includegraphics[width=.77\linewidth]{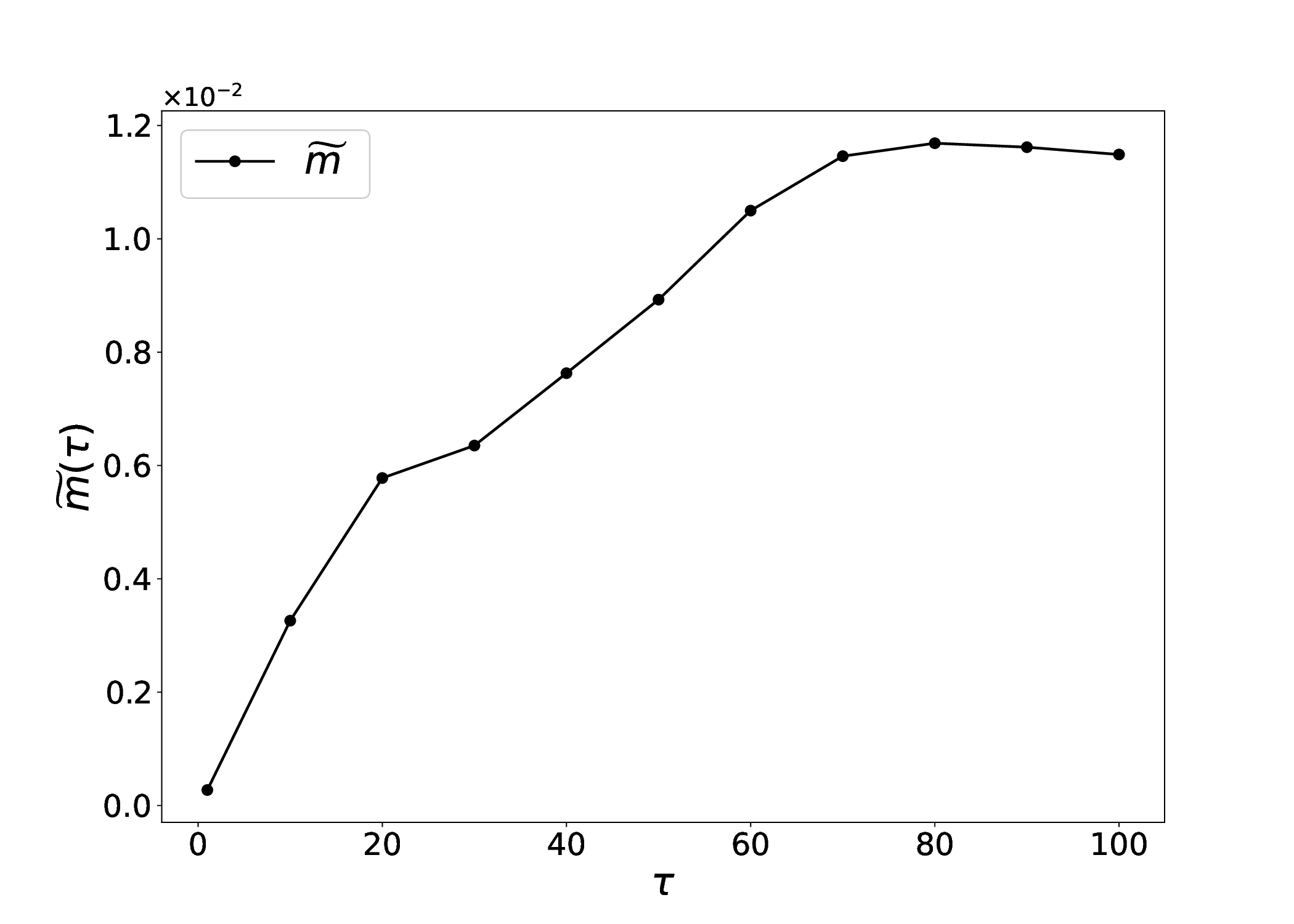}
    \caption{Median of the distribution of returns as a function of the number of days of accumulation, $\widetilde{m}(\tau)$}
    \label{mtilde}
\end{figure}

\begin{figure}[!htb]
    \centering
    \includegraphics[width=.77\linewidth]{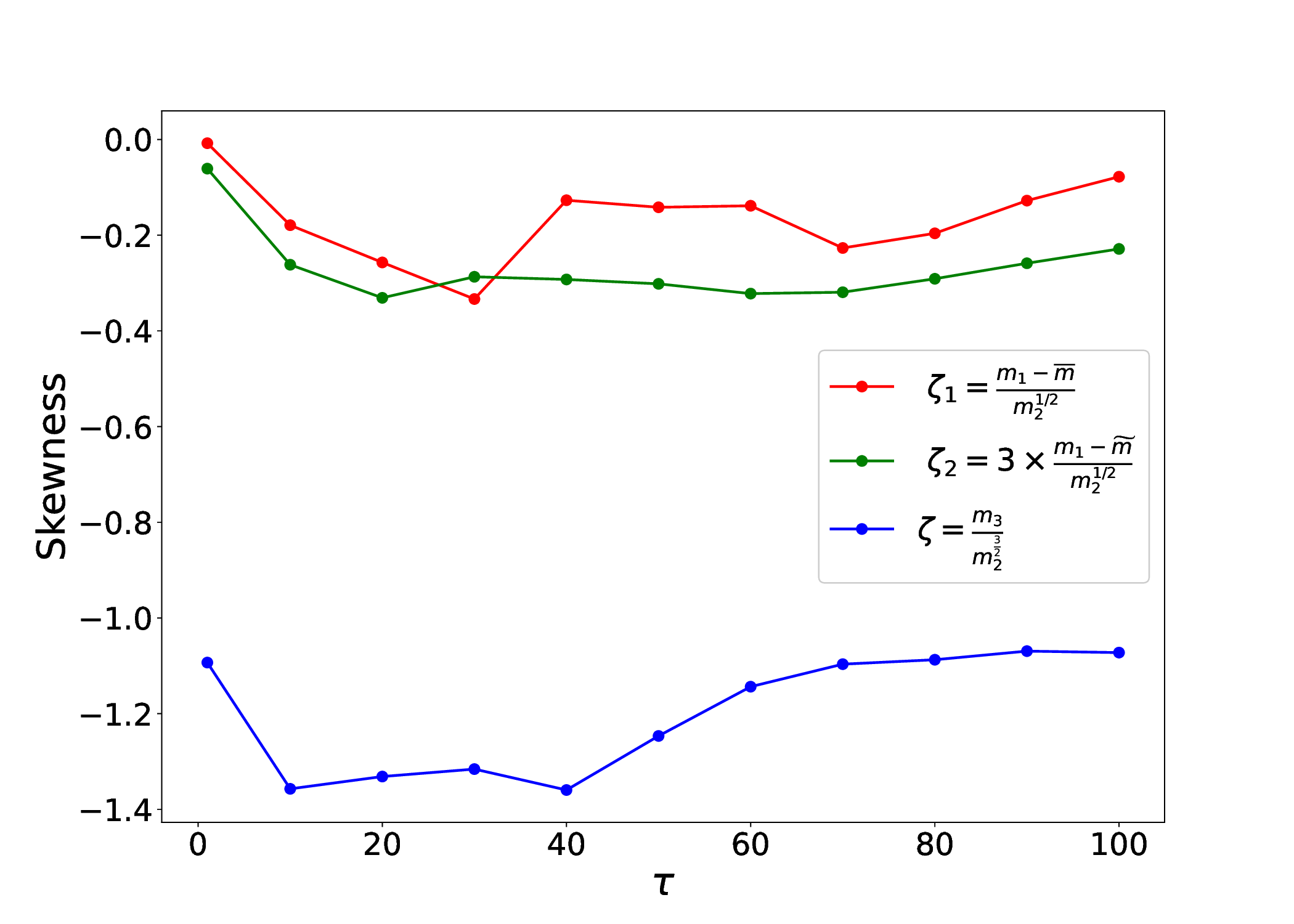}
    \caption{Fisher-Pearson coefficient of skewness $\zeta$, eq. (\ref{zeta}), and first and second Pearson coefficients of skewness $\zeta_1$, eq. (\ref{zeta1}) and $\zeta_2$, eq. (\ref{zeta2}), as function of $\tau$.}
    \label{skew}
\end{figure}

Table \ref{stats} summarizes statistics of the distributions of returns as a functions of $\tau$. Visually the dependence on $\tau$ is shown in Figs. \ref{m1} - \ref{skew}. Namely, Fig. \ref{m1} shows the mean $m_1$ of the distributions as a function $\tau$, with the linear fit and its scaled value $m_1(\tau)/\tau$. Clearly, linearity of $m_1(\tau)$ is quite good. We also point out that $m_1$ is positive and that the slope of the linear fit, per insert in Fig. \ref{m1}, is about an order of magnitude smaller than that of $\mu_{\tau}$ in Fig. \ref{mu_tau}, which is used to de-trend linear dependences in Fig. \ref{trend}.

Fig. \ref{m2} shows $m_2(\tau)$, which is nearly perfectly linear, and its scaled value $m_2(\tau)/\tau$. Significance of this linearity will be discussed in detail in Sec. \ref{theory}. Figs. \ref{mbar} and \ref{mtilde} show that the $\overline{m}$ and $\widetilde{m}$ do not have a "clean" dependence on on $\tau$, which is not surprising given the statistical definition of those quantities. Finally, Fig. \ref{skew} shows skewness $\zeta$, $\zeta_1$ and $\zeta_2$ per eqs. (\ref{zeta}) - (\ref{zeta2}). 

\begin{table}[ht] 
\centering
\begin{tabular}{|c|c|c|c|c|c|c|c|}
\hline
\textbf{$\tau$} & $m_1$  & $\overline{m}$ & $\widetilde{m}$ & $m_2$ & $\zeta$ & $\zeta_1$ & $\zeta_2$ \\ \hline
1  & $4.38 \times 10^{-5}$  & $1.32 \times 10^{-4}$ & $2.73 \times 10^{-4}$  & $1.28 \times 10^{-4}$  & -1.093  & -0.0078 & -0.0609 \\ \hline
10  & $4.21 \times 10^{-4}$  & $6.25 \times 10^{-3}$ & $3.26 \times 10^{-3}$  & $1.06 \times 10^{-3}$  & -1.357  & -0.1791 & -0.2617 \\ \hline
20  & $7.72 \times 10^{-4}$  & $1.24 \times 10^{-2}$ & $5.78 \times 10^{-3}$  & $2.06 \times 10^{-3}$  & -1.331  & -0.2569 & -0.3309 \\ \hline
30  & $1.13 \times 10^{-3}$  & $1.93 \times 10^{-2}$ & $6.35 \times 10^{-3}$  & $2.99 \times 10^{-3}$  & -1.316  & -0.3334 & -0.2869 \\ \hline
40  & $1.51 \times 10^{-3}$  & $9.48 \times 10^{-3}$ & $7.63 \times 10^{-3}$  & $3.94 \times 10^{-3}$  & -1.360  & -0.1269 & -0.2925 \\ \hline
50  & $1.92 \times 10^{-3}$  & $1.18 \times 10^{-2}$ & $8.93 \times 10^{-3}$  & $4.85 \times 10^{-3}$  & -1.247  & -0.1416 & -0.3017 \\ \hline
60  & $2.39 \times 10^{-3}$  & $1.28 \times 10^{-2}$ & $1.05 \times 10^{-2}$  & $5.70 \times 10^{-3}$  & -1.144  & -0.1385 & -0.3221 \\ \hline
70  & $2.86 \times 10^{-3}$  & $2.12 \times 10^{-2}$ & $1.15 \times 10^{-2}$  & $6.53 \times 10^{-3}$  & -1.097  & -0.2268 & -0.3192 \\ \hline
80  & $3.32 \times 10^{-3}$  & $2.02 \times 10^{-2}$ & $1.17 \times 10^{-2}$  & $7.44 \times 10^{-3}$  & -1.087  & -0.1961 & -0.2911 \\ \hline
90 & $3.72 \times 10^{-3}$  & $1.54 \times 10^{-2}$ & $1.16 \times 10^{-2}$  & $8.39 \times 10^{-3}$  & -1.069  & -0.1277 & -0.2586 \\ \hline
100 & $4.12 \times 10^{-3}$  & $1.16 \times 10^{-2}$ & $1.15 \times 10^{-2}$  & $9.33 \times 10^{-3}$  & -1.072  & -0.0778 & -0.2287 \\ \hline
\end{tabular}
\caption{Statistical Analysis of Distributions of Returns}
\label{stats}
\end{table}

\pagebreak
\section{Theoretical Framework \label{theory}}

Theoretical framework for calculating stock returns consists of two key elements. The first one is the stochastic equation for returns
\begin{equation}
\mathrm{d}x_t = \sigma_t\mathrm{d}W_t^{(1)}
\label{dxt}
\end{equation}
where $\mathrm{d}W_t^{(1)}$ is the normally distributed Wiener process, $\mathrm{d}W_t^{(1)} \sim \mathrm{N(}0,\, \mathrm{d}t \mathrm{)}$, and $\sigma_t$ is the stochastic volatility. Physical meaning of (\ref{dxt}) is that the fluctuations around the straight-line trend in Fig. \ref{trend} are due to stochastic volatility. For a known distribution of $\sigma_t$ the steady-state distribution of returns is then computed as the product distribution of $\sigma_t$ and $\mathrm{d}W_t^{(1)}$ \cite{ma2014model,liu2019distributions,dashti2021combined}.

The two widely used mean-reverting models of stochastic volatility $\sigma_t$, expressed in terms of stochastic variance $v_t=\sigma_t^2$, are multiplicative (MM) \cite{praetz1972distribution,nelson1990arch,fuentes2009universal,ma2014model,liu2019distributions}
\begin{equation}
\mathrm{d}v_t = -\gamma(v_t - \theta)\mathrm{d}t + \kappa_M v_t\mathrm{d}W_t^{(2)}
\label{IGa}
\end{equation}
and Cox-Ingersoll-Ross/Heston (HM) \cite{cox1985theory,heston1993closed,dragulescu2002probability}
\begin{equation}
\mathrm{d}v_t = -\gamma(v_t - \theta)\mathrm{d}t + \kappa_H \sqrt{v_t}\mathrm{d}W_t^{(2)}
\label{Ga}
\end{equation}
The two can be combined as follows into a MHM model \cite{dashti2021combined} (with further generalization found in \cite{liu2023rethinking})
\begin{equation}
\mathrm{d}v_t = -\gamma(v_t - \theta)\mathrm{d}t + \sqrt{\kappa_M^2 v_t^2 + \kappa_H^2 v_t }\mathrm{d}W_t^{(2)}
\label{sdrGaGa}
\end{equation}
Here $\mathrm{d}W_t^{(2)}$ is the normally distributed Wiener process, $\mathrm{d}W_t^{(2)} \sim \mathrm{N(}0,\, \mathrm{d}t \mathrm{)}$. Unless one considers such effects as leverage \cite{perello2002stochastic,dashti2021distributions} correlations between $\mathrm{d}W_t^{(1)}$ and $\mathrm{d}W_t^{(2)}$ can be neglected. 

The reason for using mean-reverting models is due to thee assumption that volatility reverts to its mean value, manifested by the fact that the mean of ${v_t}$ is given by
\begin{equation}
\overline{v_t} = \theta
\label{theta}
\end{equation}
with the implication that the mean realized volatility for $\tau$-days accumulations is given by 
\begin{equation}
\overline{V(\tau)} = \int_{0}^{\tau}\overline{(\mathrm{d}x_t)^2} = 
\int_{0}^{\tau}\overline{\sigma_t^2}\mathrm{d}t = \int_{0}^{\tau}\overline{v_t}\mathrm{d}t =\theta\tau
\label{RV}
\end{equation}
Clearly, since $\sigma_t=\sqrt{v_t}$ is a positive quantity and $\mathrm{d}W_t^{(1)}$ is even, (\ref{dxt}) produces symmetric distribution, which is contrary to what we are seeing empirically. For a symmetric distribution however we would have 
\begin{equation}
m_2(\tau)=\overline{V(\tau)}=\theta\tau
\label{m2RV}
\end{equation}
which is confirmed by a direct calculation. 

Namely, omitting HM since it does not produce power-law tails, and referring specifics to refs. \cite{liu2019distributions,dashti2021combined}, we find that the MM steady-state distribution for $v_t$ is expressed in terms of Inverse Gamma function as 
\begin{equation}
\mathrm{IGa(}v_t;\, \frac{\alpha }{\theta}+1,\, \alpha \mathrm{)}
\label{IGa}
\end{equation}
where
\begin{equation}
\alpha = \frac{2\gamma \theta}{\kappa_{M}^2}
\label{alpha}
\end{equation}
and the MHM steady-state distribution is a Beta Prime (Beta2) distribution, 
\begin{equation}
BP(v_t; p,q,\beta)=\frac{(1+\frac{v_t}{\beta})^{-p-q}(\frac{v_t}{\beta})^{-1+p}}{\beta B(p,q)}
\label{BP}
\end{equation}
where $B(p,q)$ is the beta function and the shape parameters $p$ and $q$ and the scale parameter $\beta$ are given by
\begin{equation}
\label{p,q,beta}
p=\frac{2 \gamma \theta}{\kappa_H^2}, \;
q=1+\frac{2 \gamma}{\kappa_M^2},  \;
\beta=\frac{\kappa_H^2}{\kappa_M^2}
\end{equation}
Distributions for $\sigma_t$ are then obtained by a simple change of variable $\sigma_t=\sqrt{v_t}$. As a result, the product distribution produces a Student's distribution of returns for MM    
\begin{equation}
\psi_M(z) = \frac{\Gamma \left(\frac{\alpha }{\theta 
}+\frac{3}{2}\right)}{\sqrt{\pi } \Gamma \left(\frac{\alpha }{\theta }+1\right)} 
\frac{1}{\sqrt{2 \alpha  \tau }} \left(\frac{z^2}{2 \alpha  \tau 
}+1\right)^{-\left(\frac{\alpha }{\theta }+\frac{3}{2}\right)}
\label{pdfMM}
\end{equation}
and 
\begin{equation}
\psi_{MH}(z) = \frac{\Gamma \left(q+\frac{1}{2}\right) U\left(q+\frac{1}{2}, \frac{3}{2}-p, \frac{z^2}{2 \beta  \tau}\right)}{\sqrt{2\pi \beta \tau} B \left(p,q \right)} 
\label{pdfMHM}
\end{equation}
for MHM, where $U$ is the confluent hypergeometric function. In both (\ref{pdfMM}) and (\ref{pdfMHM})  $\mathrm{d}x_t$ was replaced with $z$ and $dt$ with $\tau$. 

As mentioned above, evaluating variance using \ref{pdfMM} and \ref{pdfMHM} yields $m_2(\tau)=\theta\tau$ \cite{liu2019distributions,dashti2021combined}. Moreover, comparing values of $\theta$ found by fitting distributions of returns with (\ref{pdfMM}) and (\ref{pdfMHM}) \cite{liu2019distributions,dashti2021combined} to the values of slopes of $m_2(\tau)/\tau$ in \cite{liu2019distributions} and in Fig. \ref{m2} we observe that the two are very close (notice the difference in the values of $\tau$ and the range of years here and in \cite{liu2019distributions,dashti2021combined}).
This closeness leads to realization that the skew of the distribution of returns has little effect on the scaled second central moment $m_2(\tau)/\tau$ relative to the one expected form symmetric distributions. The latter indicates that the distribution of returns has to be only slightly asymmetric relative to (\ref{pdfMM}) and (\ref{pdfMHM}). This conclusion is also supported by a very small positive skew of the scaled mean $m_1(\tau)/\tau$.

It should be mentioned that one version of stochastic equation for returns, depending on Ito or Stratonovich interpretation \cite{dashti2021combined}, reads as follows:
\begin{equation}
\mathrm{d}x_t = -\frac{\sigma_t^2}{2}\mathrm{d}t + \sigma_t\mathrm{d}W_t^{(1)} 
=- \frac{v_t}{2}\mathrm{d}t + \sqrt{v_t}\mathrm{d}W_t^{(1)}
\label{dxt2}
\end{equation}
This interpretation produces an asymmetric distribution of returns, with negative skewness towards losses (see below). However, it does not conform to our numerical observations. To begin with, averaging (\ref{dxt2}) yields
\begin{equation}
\overline{\mathrm{d}x_t }= 
- \overline{\frac{v_t}{2}}\mathrm{d}t \rightarrow -\frac{\theta}{2} \tau
\label{dxt2av}
\end{equation}
that is a negative value of the mean. This can be "remedied" phenomenologically by replacing $\mathrm{d}x_t$ with $\mathrm{d}\left(x_t-(m_1+\theta)t\right)$, where $m_1$ is the empirically found mean above. However, there are much more substantial problems with (\ref{dxt2}). They can be traced to the fact that negative term in (\ref{dxt2}) constitutes a very small correction to the positive one until $\tau \sim 4/\theta \approx 3.8 \times 10^4$ where we used $\theta \approx .95 \times 10^{-5}$. However it is obvious from Figs. \ref{CCDF1}-\ref{CCDF20} that asymmetry between gains and losses start at much smaller values of $\tau$.

More precisely, the PDF for MM is given by \cite{liu2019distributions}
\begin{equation}
\phi_M(z) = \frac{2^{-2 \left(\frac{\alpha }{\theta }+1\right)}}{\sqrt{\pi } 
\Gamma \left(\frac{\alpha }{\theta }+1\right)}
\left(\sqrt{2 \alpha  \tau }\right)^{\frac{\alpha }{\theta }+\frac{1}{2}}
\left(\sqrt{\frac{z^2}{2 \alpha  \tau }+1}\right)^{-\left(\frac{\alpha }{\theta 
}+\frac{3}{2}\right)}
K_{\frac{\alpha }{\theta }+\frac{3}{2}}\left(\frac{\sqrt{2 \alpha  \tau }}{2} 
\sqrt{\frac{z^2}{2 \alpha  \tau }+1}\right)
e^{-\frac{z}{2}}
\label{pdfJM}
\end{equation}
where where $K$ is the modified Bessel function of the second kind of order $\frac{\alpha }{\theta }+\frac{3}{2}$. This distribution is skewed towards losses as is obvious from asymptotic behaviors:  $\phi_M(z) \approx z^{-2-\frac{\alpha}{\theta}} \left(a+b z^{-1}+c z^{-2} + ... \right)$ for $1\gg z\gg 2\alpha \tau$ and $\phi_M(z)\ \propto \exp{-z}$ for $z\gg 1$ for gains, $z>0$, and $\phi_M(z) \approx |{z}|^{-2-\frac{\alpha}{\theta}} \left(a+b |z|^{-1}+c |z|^{-2} + ... \right)$ for $|z|\gg 2\alpha \tau$ for losses, $z<0$, \cite{nist2022digital} where $a,b,c$ are functions of $\alpha$, $\tau$ and $\theta$ and  $\alpha \approx 1.5 \times 10^{-4}$ \cite{liu2019distributions}. Consequently, the distribution (\ref{pdfJM}) is symmetric until roughly $z \sim 1$ which, again, not the case per Figs. \ref{CCDF1}-\ref{CCDF20}. Finally skewness produced by fitting with (\ref{pdfJM}) yields negative skewness which is an order of magnitude smaller than that in Fig. \ref{skew}.


\section{Conclusions \label{conclusions}}

We conducted empirical analysis of S\&P500 (de-trended) daily and multi-day (accumulated) returns over the 1980-2024 time period which included the four major market calamities: Black Monday (in the midst of Savings and Loans crisis), Tech Bubble, Financial Crisis and Covid Pandemic. We examined gains and losses separately for existence of power-law tails and for outliers, such as Dragon Kings and negative Dragon Kings. The two main findings in this regard is that losses have far heavier tails than gains and that the power-law description of the tails deteriorates considerably with the number of days of accumulation. For the latter, the rapid fall-off of points in tail ends -- which correspond to the biggest gains and losses -- from the power-law dependence can be possibly characterized as negative Dragon Kings. This fact may be attributed to the breadth and strength of the S\&P500 index which prevents longer-term unrestrained gains and losses.

We also examined statistical characteristics of full distributions of returns and found that with good accuracy the mean increases linearly with the number of days of accumulation, while variance increases linearly with even higher precision. We also found that skewness of the distributions is negative -- consistent with the heavier tails of losses -- and is only weakly dependent on the number of days of accumulation. In this regard, we discussed  theoretical framework, based on a pair of stochastic differential equations for returns and stochastic volatility, for describing distributions of returns. Such approach produces either symmetric or near-symmetric distributions which are in excellent agreement with the linear dependence of the variance but fail to describe the properties of the mean and the skewness.

\newpage

\bibliography{mybib}

\end{document}